**Multiple species animal movements: network properties, disease dynamics and the impact of targeted control actions**

Nicolas C. Cardenas[1], Abagael L. Sykes[1], Francisco P. N. Lopes[2], Gustavo Machado[1,*]

[1] Department of Population Health and Pathobiology, College of Veterinary Medicine, North Carolina State University, Raleigh, North Carolina, USA.

[2] Departamento de Defesa Agropecuária, Secretaria da Agricultura, Pecuária e Desenvolvimento Rural (SEAPDR), Porto Alegre, Brazil.

Infectious diseases in livestock are well-known to infect multiple hosts and persist through a combination of within- and between-host transmission pathways. Uncertainty remains about the epidemic dynamics of diseases being introduced on farms with more than one susceptible host species. Here, we describe multi-host contact networks and elucidate the potential of disease spread through farms with multiple hosts. Four years of between-farm animal movement among all farms of a Brazilian state were described through a static and monthly snapshot of network representations. We developed a stochastic multilevel model to simulate scenarios in which infection was seeded into a single host and multi-hosts farms to quantify disease spread potential, and simulate network-based control actions used to evaluate the reduction of secondarily infected farms. We showed that the swine network was more connected than cattle and small ruminants in both the static and snapshots temporal network analysis. The small ruminant network was highly fragmented, however, contributed to interconnecting farms with other hosts acting as intermediaries throughout the networks. When a single host was initially infected, secondary infections were observed across farms with all other species. Our stochastic multi-host model demonstrated that targeting the top 3.25% of the farms ranked by degree reduced the number of secondarily infected farms. The proposed simulation results highlighted the importance in considering multi-host dynamics and contact network while designing surveillance and preparedness control strategies against pathogens known to infect multiple species.

## Introduction

Infectious diseases in livestock populations have the potential to create large outbreaks and epidemics which compromise animal health and welfare, and produce economic losses [1]. Historically, these epidemics have been associated with disease transmission among multiple livestock hosts, i.e., the 2001 foot-and-mouth disease (FMD) outbreak in the United Kingdom [1]. To better understand the main drivers of such large-scale outbreaks, several studies have used social network analysis (SNA) and mathematical models to characterize animal movement patterns, often with a focus on individual food animal species such as cattle [2–5] or swine or small ruminants [6]. Only a few studies have developed mathematical models or even described contact networks over animal movement data of multiple species at the same time [7–16].

In previous literature, social network analysis has been used to shed light on between-farm disease spread processes, including the calculation of expected epidemic size where a wide range of network features derived from both static and monthly snapshots of network representation have been implemented [2, 17, 18]. The use of such approaches is very useful to describe the topology and temporal variation in the networks, however, it can lead to overestimation of the connectivity of the animal trade. Therefore, approaches that account for the temporal order of movements such as temporal contact chains analysis and disease spread models have been successful in capturing epidemic trajectories [19–21].

The broader availability of animal movement data has been followed by significant advancements in disease spread modeling [22–24]. Those studies revealed important dynamic characteristics of epidemics that could be used in the development of targeted surveillance systems for farms with a higher likelihood of infection through animal movements [10, 25]. Our study was conducted with data from one Brazilian state. Here, we implemented analysis that allows for the description of the possible risks of animal movements, and identify key farms in the contact network using a multispecies approach. Social network analysis and mathematical models considering more than one host can deliver insights into the quantification of potential outbreak paths and provide useful data for the identification of key farms that can be used to break the direct contact transmission of animal diseases [8, 14, 26–29]. These insights are invaluable in supporting the design and implementation of efficient control and eradication strategies. From our detailed information about the entire movement data and population data we aimed at i) characterize single-host and multi-host (bovine, swine and, small ruminants) contact networks across four years of movement data; ii) examine the potential of spread disease from one specific host to a different host considering the animal movements as transmission route; and iii) measure the impact of network-based target interventions to identify key nodes that could affect the number of infected premises in each host using a stochastic multi-level susceptible-infected model.

## Material and Methods

Daily between-farm movement data from January 2015 to December 2018 of all registered farms with at least one of the following food animal hosts: cattle, buffalo, swine, sheep, and goat; were collected under a data use agreement with the Rio Grande do Sul Secretary of Livestock and Irrigation (SEAPI-RS, 2018) in Brazil. The data was extracted directly from the state's database and included 1,621,374 movement records. Incomplete or inaccurate data including movements with missing farm identification, the same movement origin, and destination, or movements across state borders (n= 16,956) were not considered for the analysis. The final database included 1,604,418 movements and 90,090,619 animals; 944 records were fairs and events in farms corresponding to 467 nodes, which represents 0.16% of the total number of premises. Movements to slaughterhouses (34.56%, n= 554,487) were not included in the network description but were used in the disease spread model. For descriptive purposes cattle and buffalo hosts were grouped as "bovine", and sheep and goat hosts were grouped as "small ruminants''.

***Static and monthly snapshots of networks representations***

The between-farm total of animals moved (referred herein as "batch") was represented as a directed bipartite graph, $g$, where each farm was represented as a "node" and the movements between farms were represented as "edges". Each edge has a specific node origin, $i$ and a specific node destination, $j$. Each host network was analyzed individually and a full network (which includes all hosts) was reconstructed using data from all four years. Furthermore, in this static network description, node-level metrics: degree and, betweenness, were described by calculating the global mean with their respective confidence intervals of 95%, according to the network-host and period considered. The monthly snapshots of the static host networks of the full contact network were constructed to assess the temporal development of the network for the whole period of study using a monthly and yearly time resolution.

For both networks, we calculated network-level metrics including the number of nodes, edges, and animals; the diameter of the network; mean of degree; betweenness; PageRank; global cluster coefficient; and centralization based on degree. In addition, we also calculated the sizes of the connected components: Giant Strongly Connected Component (GSCC), and Giant Weakly Connected Component (GWCC) (Table 1). Based on the static network, the in-degree and out-degree distributions were calculated for each species and for the full network.

***Contact chains analysis***

We used contact chain analysis to describe the temporal and sequential nodes accessible by the formation of edges over time. These chains were divided into two main types: the in-going contact chains (ICC), which identify the number of farms that could potentially transmit the infection to the index farm over a defined period arising from the purchase and importation of animals; and the out-going contact chains (OCC), which can be used to quantify the number of farms that could potentially acquire infection from the index farm through the onward sale and export of animals [19, 30].

**[Table 1]**

**Dynamic model description**

The proposed model was designed to demonstrate the potential for the spread of infectious animals from one specific host to a different host, considering only transmission through animal movements. We also measured the impact of SNA centrality metrics in a theoretical node removal in the network, also referred to as percolation of nodes, to identify key nodes that should be targeted to implement control actions. These control actions prevent farms from receiving or shipping infected animals such as culling, isolation of animals, increased hygiene measures or vaccination [20, 31–33]. So far, many past approaches are based on static networks where the temporal nature of animal movements and the vital dynamics (birth and deaths) of animals inside each farm is neglected. For that reason, we incorporated within farms and between dynamics through a Susceptible-Infectious (SI) model using the temporal animal movement data explicitly with a higher effective contact rate to ensure an efficient disease transmission over the simulations. Here, each node removed from the network cannot transmit disease to other nodes, allowing the quantification of the effectiveness of each strategy by the number of secondarily infected farms at the end of the simulations.

### ***Within-farm dynamics***

We implemented a discrete dynamic stochastic model that accounts for: i) the local dynamics that represent each animal population (bovine, swine, and small ruminants); and ii) the global dynamics that represent the interaction within farms and slaughterhouses; using a time resolution of one day. This transmission model framework is implemented using the *SimInf R* package for data-driven stochastic disease spread simulations [34] and the structure is described below.

Here, we tailored the within-farm dynamics to account for state transition from susceptible, $S$, to infected, $I$, compartments at a rate proportional to a frequency-dependent transmission coefficient, $\beta = 0.7$ in the simulated population. This assumed a homogeneously mixing population, $i$, within each premise where all hosts had identical rates of infection. The model considers the between-farm movements among farms of all species, therefore considering a real multi-host contact network of movement data from January 1st to December 31st, 2018. Of note, we did not explicitly model the heterogeneities related to animal space and geolocation, behavioral aspects, or animal categories such as age group for bovine or farm type in swine farms.

The initial simulation procedure was seeded in 1,000 farms, each simulation. This was repeated so that each single-host farm or multi-host farm was the initial seeded farm for a total of 100 simulations. Transitions of animals between the susceptible and infected compartments were modeled as a continuous-time discrete-state using Gillespie's direct method continuous-time Markov chain [34], as follows,

$$\frac{dI}{dt} = \frac{\beta S_i I_i}{S_i + I_i}, \tag{1}$$

where $t$ represents a specific time in a farm $i$.

***Between-farm dynamics***

For the between-farm dynamics, susceptible farms, $S_i$, can be infected by their contacts via daily animal movements. Within these dynamics, animal movements to a slaughterhouse destination were considered to be subtracting these animals from the system. The number of eligible animals moved between nodes and slaughterhouses were selected according to the daily movement records for each host. Each draw of animals from $S$ and $I$ compartments were at random, while the total number of available animals were based on the official data provided by [50] for each premises. To assure the stability of the farm populations, the animals born alive and declared to official veterinary service were used as an "enter event" and were assumed to be susceptible hosts. We subsequently calculated the number of ingoing and outgoing animals and the number of animals born and deceased for the full period of simulation, to update the animal population in each farm. For any site that have ran out of animals we reset the original number of animals reported to state database [50]; this allows us to avoid sampling animals from non-existent populations. In Figure 1, we depict the interaction of the multiscale model considering both within-farm and between-farm dynamics. At the end of each simulation, the daily number of secondarily infected farms of each host was calculated.

[**Figure 1 here**]

***Simulating network-based targeted percolation***

Our analysis utilizes a theoretical percolation of farms (nodes) to remove nodes and their edges from the network [20, 35, 36]. Thus, these nodes are removed from the contact network as are all their associated animal movements, leading to a fragmented network that allows us to identify important farms whose removal reduces secondary infections.

We additionally considered the effect of sequential removal of farms from the network, based on their ranked order. Here, we used the static network constructed with movement data from 2015 to 2016, and the full network parameters of degree betweenness and PageRank (Table 1) to rank farms in descending order. We then quantify the effect of sequential removal on the reduction of secondarily infected farms in the host populations. Furthermore, a sample of nodes was generated to be removed to simulate the non-targeted implementation of control actions from 2017 to 2018. The targeted farms were removed before model simulation was started to fragment the contact network, and the number of infected farms was then calculated at the end of each full simulation, on day 730 (December 31th 2018). For each simulation, the number of secondarily infected farms was grouped by host species and network metric, and utilized in the identification of the key nodes to target with control actions. In this analysis, the "without control" scenario is represented as having zero nodes removed from the network since it represents the total number of infected farms observed when no control actions have been implemented. In the subsequent "with control" scenarios, all the simulations started with values taken from the end of the "without control" scenario, after which $n$ nodes (ranging from 1 to 10,000) were removed and the impact over the number of secondary cases measured.

***Initial conditions and farm selection***

Since one of our objectives was to quantify the potential of a given infected host infecting another susceptible host(s) via animal transport, we developed scenarios A to C where the infection was seeded at bovine, small ruminants, and swine single host farms, respectively. Transportation of infected animals across multi-host farms was the only mode of disease propagation into other host species sites. In scenario D infections were seeded in multi-host farms with at least two species (e.g. bovine and swine) and in scenario E farms were randomly selected from the full network and seeded (Table 2). Each simulation started with 1,000 infected farms drawn randomly from the farm list as the initially-seeded infections according to the model scenario (Table 2) and repeated 1,000 times. In each initial infected farm, 10% of the animal population was assumed to be infected. Finally, the distribution of the number of secondarily infected farms at 5, 10, 15, and 30 days of the simulation was analyzed and presented according to the proposed list of scenarios (Table 2).

For the network-based targeted percolation analysis, we created three scenarios: i) infection starting in bovine farms; ii) infection starting at swine farms, and iii) infection starting at small ruminant farms. For each scenario, we initiated each simulation infecting 1,000 random farms with a within-farm prevalence of 10 % similar to the procedure described above. In each proposed scenario, we ran the model without removing any node for 100 simulations and then quantified the mean and the confidence interval of the number of secondarily infected farms. We then selected one farm to be removed from the full network, according to a farm-ranked list based on a specific network metric and again ran 100 simulations to calculate the mean and the confidence interval of the number of secondarily infected farms. We repeated this procedure increasing the number of removed farms by one until 10,000 farms were removed. Therefore, each scenario resulted in over a total of 1 million unique and individual simulations which are independent of other simulations.

**[Table 2 here]**

**Results**

The total number of between-farm cattle and buffalo (bovine) movements were 1,251,615 (78.01%) batches and 20,773,516 (24.3%) animals, swine were 273,223 (17.02%) batches and 62,292,763 (72.8%) animals, and small ruminants represented 96,398 (6%) batches and 2,731,304 (2.89%) animals.

Bovine farms had the highest number of farms in the network followed by small ruminants and swine. For multi-host farms, the largest proportion was bovine and small ruminants with 10%, swine and bovine with 3.13%, all three host species with 0.37%, and swine and small ruminants with 0.04% respectively (Figure 2A). Overall for the between-farm movements, bovine-to-bovine (41.5%) movements were the highest, however, 27.6% of small ruminants movements were between farms with bovine, followed by swine-bovine movements with 12.83%, swine-swine movements with 8.42%, small ruminants-to-small ruminants movements 7.95% and small ruminants-to-swine movements with 1.7% (Figure 2B and Additional file 1 in Table 1). The density of these farms by the municipality is presented in the additional file 1 Figure 1.

[**Figure 2 here**]

### ***Static network description***

As expected, bovine networks had the largest number of edges (Table 3). Mean betweenness was dominated by bovine and swine farms, while the clustering coefficient remained similar among all host species and centralization was significantly higher among swine farms (Table 3). When we compared the size of GWCC and GSCC, bovine farms had the largest connected components followed by small ruminants and swine. Similar patterns were found in the yearly temporal window presented in Supplementary file 1 Table 3.

**[Table 3 here]**

### ***Distribution of in- and out-degree per host***

In Figure 3, in- and out-degree distributions are presented individually for each host network. In general, for the full network, in-degree was higher than out-degree for bovine and small ruminants, while the swine network results indicate that after a degree value of 34 the frequency of out-degree was higher than in-degree.

[**Figure 3 here**]

### ***Monthly snapshots of networks representations***

The monthly network representations are shown in Figure 4. The movements between swine farms have the highest mean degree, while bovine dominate in nodes, edges, and diameter throughout the four years. As expected the swine monthly-snapshot network exhibited a higher number of animals per batch than the other species. The calculated mean betweenness showed three unusual peaks for the bovine monthly-snapshot network; these are related to 49 nodes all associated with fairs, events, and exportation ports of bovine exportation, each one with betweenness values greater than 100,000 during the peak months, presented in supplementary Figure 4. Small ruminants networks were steady throughout the four years. When evaluating the cluster coefficients, bovine and swine movements remain steady while small ruminants have large fluctuations. The swine network centralization was much higher than bovine or small ruminants, which could be related to how the industry is organized with the formation of groups of producers with commercial contracts with an integrator [21, 37]. For the GSCC and GWCC, the bovine monthly-snapshot network followed the same temporal patterns with the nodes indicating an important dependency in the number of active farms, while small ruminants showed a seasonality with a higher percentage of GSCC and GWCC at the end of each year. Finally, the swine monthly-snapshot network exhibited the highest values of GSCC and GWCC, indicating it was the most connected network.

[**Figure 4 here**]

***Contact chains***

The out-going contact chain (OCC) of the full network and the bovine network had similar distributions ranging from 10 to 10,000 farms. The OCC distribution for small ruminants showed a lower interquartile range when compared with the other distributions, while swine OCC showed the largest interquartile range with higher densities in values above 100 farms (Figure 5).

The in-going contact chain (ICC) of bovine, swine, and small ruminant distributions follows a similar pattern of OCC counterparts, however, with lower distribution sizes compared to their OCC. Bovine and small ruminants follow the same pattern, with values clustered between $10^0$ to $10^2$. However, the contact chains for swine present very differently for both ICC and OCC compared to other hosts, with a wider interquartile range indicating larger chains to a wider range of farms. The highest density in the swine OCC was centered within 100 and 900 values, indicating a larger number of connected farms within that range (Figure 5).

[**Figure 5 here**]

[**Figure 6 here**]

In addition, we explore the monthly ICC and OCC distribution by the host. The result showed similar distributions between the full network and the bovine network. However, the swine contact chains presented higher distributions than the other species, while the small ruminants showed the lowest values but with a temporal trend with higher values distributions in December and January. All results are presented in Supplementary Figures 1 and 2.

***Assessment of secondary cases by host***

Simulations were divided by their initial seed infection conditions according to the model scenarios A, B, C and D. The estimated secondary cases, represented by the number of farms for the first 5, 10, 15, and 30 days of the disease propagation, are shown in Figure 7. For all simulated scenarios, secondary infections were observed across all host species. As expected, the highest number of infected farms corresponds to the species where the infection began due to most of the connections being intra-host. However, the connections between different hosts across multi-host farms made disease propagation to other hosts possible. It is important to note that large variabilities were observed in the distributions of swine farms followed by small ruminants and bovine, respectively. Furthermore, when simulations of the infection started in swine or small ruminants farms, there were no secondary infections between 5 to 30 days, which could be due to saturation of the current animal movement connections. A longer duration of time would be needed for farms to create movement connections with new farms.

[**Figure 7 here**]

The results are also presented as prevalence in supplementary Figure 5, to allow for a comparison of the prevalence between each host. The prevalence is proportional to the number of farms per host, as the prevalence corresponds to the number of infected host farms divided by the number of farms which had at least one host species. Under these model considerations, swine farms were the most infected species when simulations started in multi-host farms. Moreover, the prevalence of small ruminants was higher than bovine for the majority of simulations with the exception of simulations that started at bovine farms. For bovine farms, when the simulation began in any other species, results showed the lowest prevalence of all hosts (Supplementary Figure 5). Additionally, in Supplementary Table 2, we show the descriptive statistics for the farm prevalence in the 5, 10, 15, 30, days of spread disease simulation.

***Effectiveness of network-based targeted percolation***

Based on the simulation and subsequent theoretical removal of farms in the full network, the most effective network centrality metric throughout all hosts was the degree, followed by betweenness and PageRank, respectively. In Figure 8, the area of each color represents the number of infected farms on the last simulation day (December 31th, 2018) after performing the control action, which involved selecting $n$ number of farms to be removed according to their ranked network metrics. The nodes were removed from the total population, and subsequently, the prevalence in each host was quantified. For example, in the swine panel, Figure 8, it is evident that random node removal from the total farm population is not efficient to break the full network and in consequence, could not reduce the swine farms prevalence since the removed nodes are not related to the pig farms' connections. Whilst, the removal of 1,000 farms ranked by degree was sufficient to reduce prevalence values close to zero, while for small ruminants and bovine it was not possible to reach close to zero infected farms, even with 10,000 farms removed from the networks. For the bovine farms, the removal of 10,000 farms induced a reduction in prevalence from 5% to 1% (Figure 8 on the right y-axis). The list of the top 10,000 farms ranked by degree includes 5,316 bovine farms, 4,080 multi-host farms, 418 swine farms, and 43 small ruminant farms.

[**Figure 8 here**]

**Discussion**

In this study, we described between-farm contact networks of individual host species and a full network considering movements among bovine, small ruminants, and swine farms in one Brazilian state. Using the full network we developed a compartmental network model to assess the potential for disease spread through the contact networks while considering the implications of network-based control actions. Based on our descriptive analysis, we found higher connectivity in the static bovine network, while in the temporal and monthly snapshot network representations analysis we found higher connectivity in the swine network. In contrast, small ruminants farms were found to be widely disconnected in all network views, monthly snapshot static networks representations, and contact chains analysis approaches. In the Susceptible-Infected stochastic multi-host model, where we seeded infection individually to each host species, results demonstrated that transmission to other host-species was not dependent on the initial seeded species. In addition, we simulated node-removal from the full network, by ranking farms based on their network measure (i.e. degree) and subsequently removing the first 10,000 farms, showing that ranking by degree would reduce the simulated prevalence by from 5% to 1% (Figure 8). Thus, we argue that it is sensible for disease control and surveillance programs for pathogens capable of infecting multiple hosts, to consider the usual movement restrictions during disease emergencies, but also to consider surveillance activities such as serosurveillance at all susceptible species, this is particularly relevant for foreign animal disease preparedness activities against FMD for example.

In the static network view, bovine networks accounted for 80.97% of the total movements and constituted 94.54% of farms in the database, generating the largest giant connected component sizes (GSCC and GWCC) and network diameter (Table 2). Within the full network, bovine farms acted as hubs that influenced the network dynamics by creating the presence of possible super spreader nodes such as event farms with a high out-degree, which has been described in other studies performed in Brazil [3–5]. In the current study, fairs and events movements represented only 0.16% of the total volume of all transported animals, which differed greatly from movement patterns in other countries such as in the UK where 30.4 % of movements were related to movements to markets [38]. In the monthly snapshot network representations view, the swine network was more connected than all other host species, with higher values of animal movements, mean of degree, centralization, and giant connected component percentages. This is likely because of the pyramidal structure commercial pig production utilizes to improve performance and productivity. Because of such continuous flow, nurseries and wean to finishers tend to receive and send a great number of pigs, thus have a higher degree in comparison with other production types [21, 37, 39]. In contrast, the network connectivity in the bovine farms was low but with more than 20,000 edges and node interaction in a monthly view representing the majority of the full network dynamics. This result was noticeable by the distribution of the degree parameter, in which all hosts exhibited power-law distributions, suggesting that these networks are characterized as scale-free [19, 40]. Additionally, the degree distribution results could help highlight super-spreader farms in all host populations through the identification of farms with several commercial partners that make a major contribution to disease spread [41]. In the case of the swine network, as expected was the network with the largest degree, which has been described previously in the same region [2, 21]. Finally, the analysis of the small ruminants' network showed a very disconnected network with a marked seasonality at the end of each year but with fewer monthly edges and nodes than swine and cattle farms, despite having more registered farms than the swine production in the full network.

Our modeling results showed that regardless of which species was first infected at the beginning of the simulations, secondary infections were detected in all farm types. This indeed is likely associated with the network structure and how food animal populations and production are organized in the study region. However, to the author's knowledge, this is the first in-depth characterization of a multi-host animal network in Brazil. We argue that uncovering the interconnection among multiple species is of great relevance because most disease control programs are centered in only a single species, often neglecting the potential of disease introduction from a different host species (e.g. FMD, influenza).

When infection started in multi-host farms the distribution of the final number of infected farms at the end of the simulation, bovine farms had a higher median than the other species, while the swine farms showed broader distributions compared to small ruminants and bovines. This may be explained mainly by how commercial pig production is organized, often extremely vertically integrated with high volume of pigs moving from farm-to-farm based on age/production phases. Finally, in all scenarios with simulations seeded in a single host (i.e. swine) the estimated number of infected farms at 30 days post seeded infection was higher than the simulations which seeded infection in the multi-host farms. This is due to the fact that the connectivity of single host farm networks allows for the spread of disease quickly, infecting more farms of the specific host species in the 30-day time frame than when the disease is started in a multi-host farm. Similar patterns have been observed in past epidemics, such as the FMD [14, 15, 26, 42, 43] where multi-host farms had a role in infection propagation.

We used our proposed model to also simulate network-based interventions, thus we used node-removal strategies to test control actions. Targeting and removing farms sequentially based on network metrics, was most impactful when the degree was used. This reduced the expected number of infected farms from ~ 5% to ~ 1% in the bovine network (degree), 0.6% to ~ 0.01 in the swine network (degree), and ~ 1 to ~ 0.025% in the small ruminants network (degree). In contrast, random node removal was the least efficient in reducing secondary infections. An unexpected result was that for small ruminants random removal of nodes had a better impact in reducing transmission than using PageRank network metric, partly because this network is more disconnected compared to the other host-networks which leads to PageRank failing to recognize important farms for the spread of disease. Previous studies that also utilized large datasets and similar node-removal approaches but single host networks, reported similar results as ours, especially for bovine networks [33, 35, 44, 45] and swine networks [21, 37, 41]. Studies that considered more than one host species and network-based target control actions reported a larger reduction in the expected number of cases directly generated by one case in a population , size of the connected components, and the number of infected farms using degree-based interventions [10, 14]. An important limitation of the studies listed above is the use of static networks which could have overestimated the connectivity of those networks [14, 20, 45]. Therefore, we argue for the need to consider temporal models when developing network-based target approaches, because it considers the temporal order of the connections and avoids the over connectivity of static networks [20, 45]. Furthermore, network analysis approaches have been widely used recently; the great majority centered on describing the contact networks of a single species, here we cite just some relevant examples [2, 18, 32, 39, 40, 46–48], however, only a fraction of studies have explicitly considered the interaction of possible contacts among more than one species while constructing their transmission networks [7–9, 11–13, 15, 16, 49]. While single host networks are informative, they are likely to underestimate the epidemic propagation of pathogens capable of infecting multiple species [26, 50–53]. Our results reinforce the relevance of including data, when available, of all susceptible host species while developing regional disease preparedness activities into disease surveillance systems and agree with previous literature which reported better performance of temporal network measures in identifying farms that could be targeted to reduce the secondary number of infections [39].

**Limitations and further remarks**

Our main goal in this study was to quantify the role of multi-host farms in disease transmission via contact networks and identify hubs in the network. Therefore, while we are aware that fomites such as delivery trucks and local-area spread are important for disease propagation [50, 54], we have not included such transmission pathways in our model nor have we included the heterogeneity present in disease transmission for any specific disease. However, in future work, we will specifically set transmission coefficients for each host species (e.g. swine, cattle) interaction, and incorporate spatial transmission within a full network model to model for example the spread of FMD. Additionally, we did not consider animal level dynamics (i.e. age, commercial versus backyard swine farms) so the risk each animal posed for disease transmission was assumed to be the same, which could contribute to an overestimation of the transmission potential. Similarly, this is not an individual-based model so we cannot individualize the results for a specific individual or calculate the time that an animal spends on each farm. Furthermore, we are unable to quantify the effect of other between-farm movements such as veterinarians and employees, which may contribute to the transmission through indirect contacts.

**Conclusion**

Using a stochastic network model, our results demonstrate that outbreaks starting in a particular host species resulted in outbreaks on farms with all other host species. Small ruminant farms, despite the lower number of nodes in the network, share a large number of contacts among other host-nodes in the full network. In addition, our multi-host network-based intervention model showed that identifying and deploying surveillance to the 10,000 farms with the largest degree, regardless of host-species, would reduce more than 70% of the simulated transmission. Ultimately, this study helps to understand the temporal variation in between-farm movement fluctuations and the topology of each contact network, as well as the similarities and differences among the three species' movement dynamics.

**List of abbreviations**

FMD: foot and mouth disease; OVS: official veterinary service; SNA: Social Network analysis; GSCC: giant strongly connected components; GWCC: giant weakly connected components; ICC: in-going contact chains; OCC: out-going contact chains

**Figure Legends**

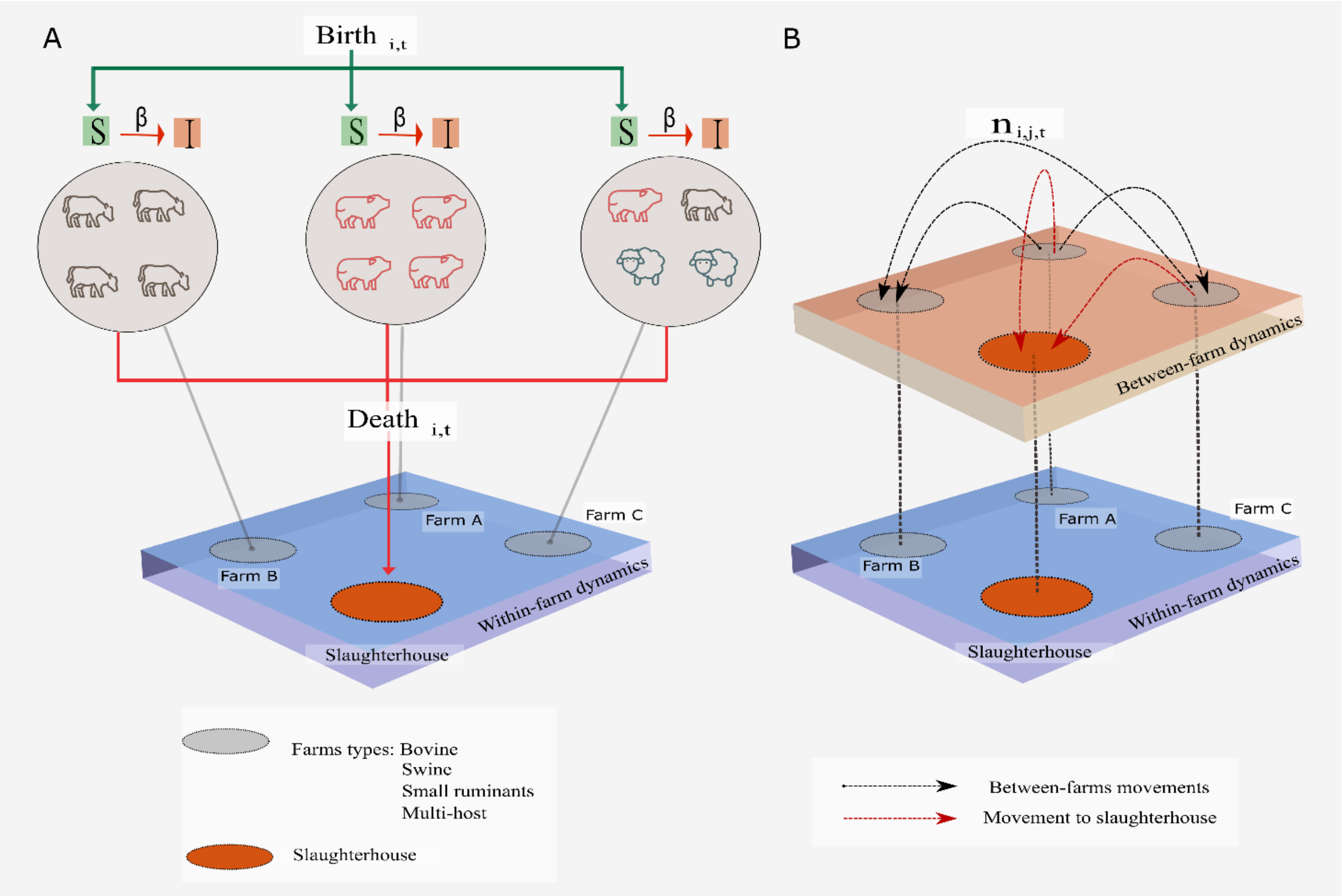

**Figure 1. Schematic of state transitions for between within-farm and between-farm**. A) Green lines indicate introduction events (births) in a specific farm, $i$, at a specific time, $t$, into a susceptible compartment, *S)*. Within-farm dynamics represent farms types with individual hosts (bovine, swine, and small ruminants) modeled in the stochastic compartmental simulations that classify into Susceptible, *S*, to Infected, *I*, states according to the transmission coefficient $\beta$. The animals that were moved to the slaughterhouse are represented as deaths on the farm, $i$, at the time, $t$, and are indicated by red lines. These animals were randomly chosen and removed from the simulation. B) The between farms dynamics layer represents the number of animals moved (batch), $n$, from the origin farm, $i$, to a destination farm, $j$, in a specific time, $t$, (indicated by the black arrows). The red arrows represent the unidirectional movements of animals from farms to slaughterhouses. Susceptible individuals were assumed to have never been infected given the absence of any foreign animal diseases within the study area for more than 10 years [55, 56].

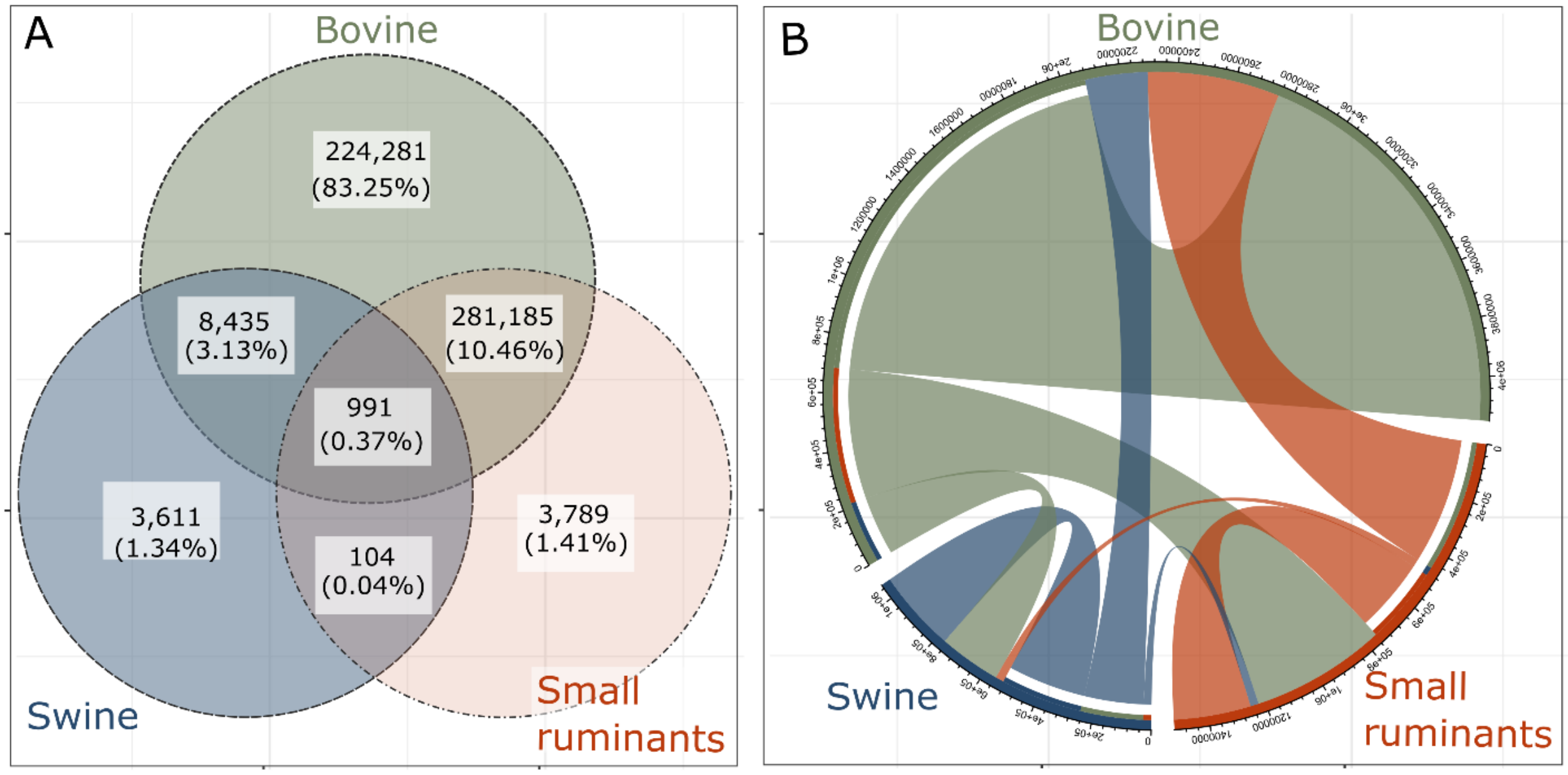

**Figure. 2 Farm population and movements among hosts.** A). Venn diagram of the number of farms and the proportion of farms among the three host groups. B) Circular plot for intra- and inter-host movement flows. Each external sector of the circle represents the origin of the movements where the numbers represent the total of the ongoing and out-going movements for each host according to the number of interactions. The out-going animal flow starts from the base of each sector.

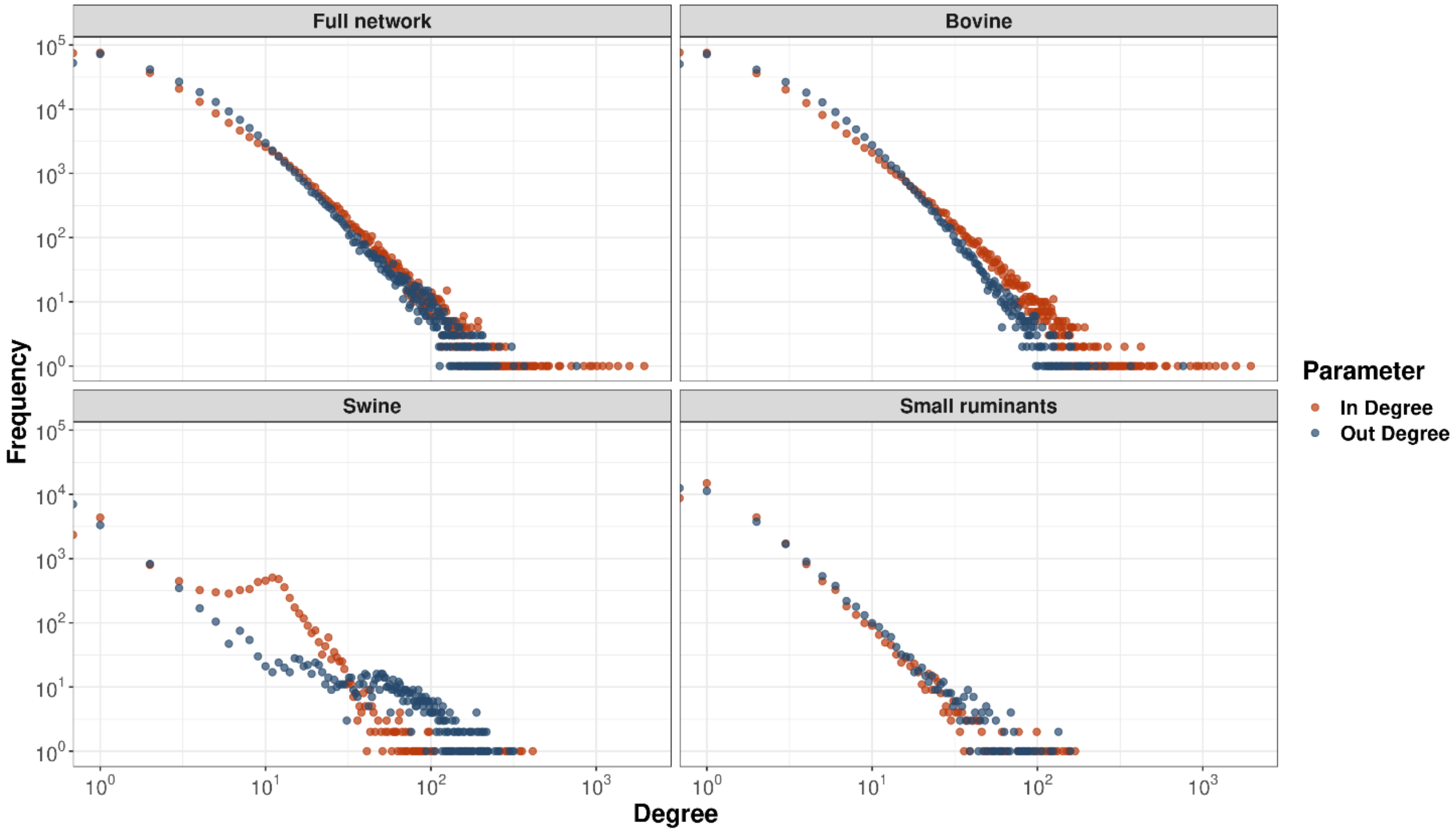

**Figure 3. Distribution of in- and out-degree data by host and for the entire network 2015-2018**. Each point represents binned data for frequencies and degree values.

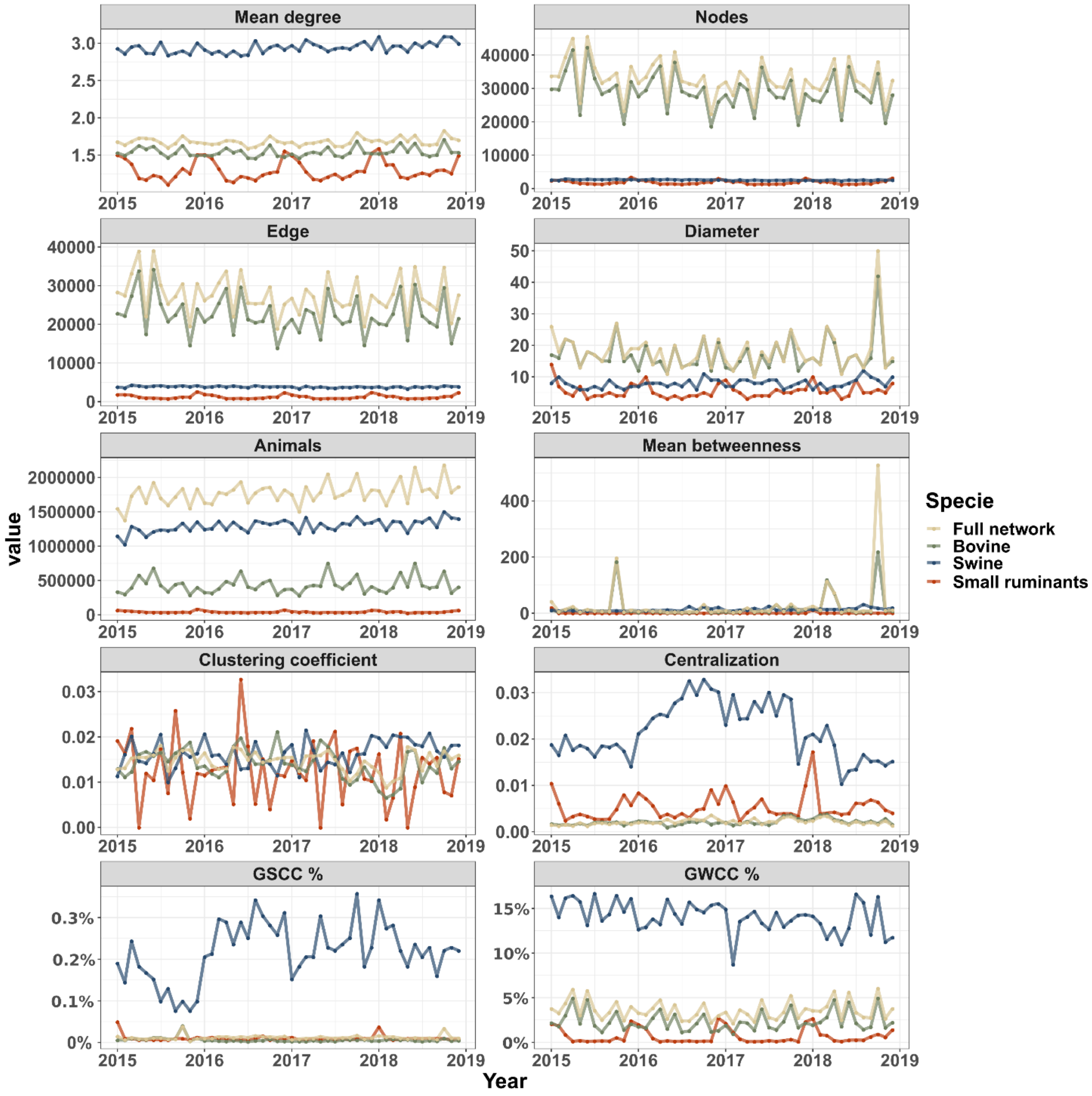

**Figure 4. Descriptive statistics of time series of static networks by the host from 2015-2018.** Mean degree is the average degree calculated by measuring the mean degree of all farms in the selected population, nodes is the number of active farms, edges are the sum of the between-farm movements (batches), animals the sum of the transported host, diameter is a number of steps in the monthly network, mean betweenness mean of the node-shortest paths, cluster coefficient and centralization the proportions and, GWCC and GSCC the percentage of farms in each component.

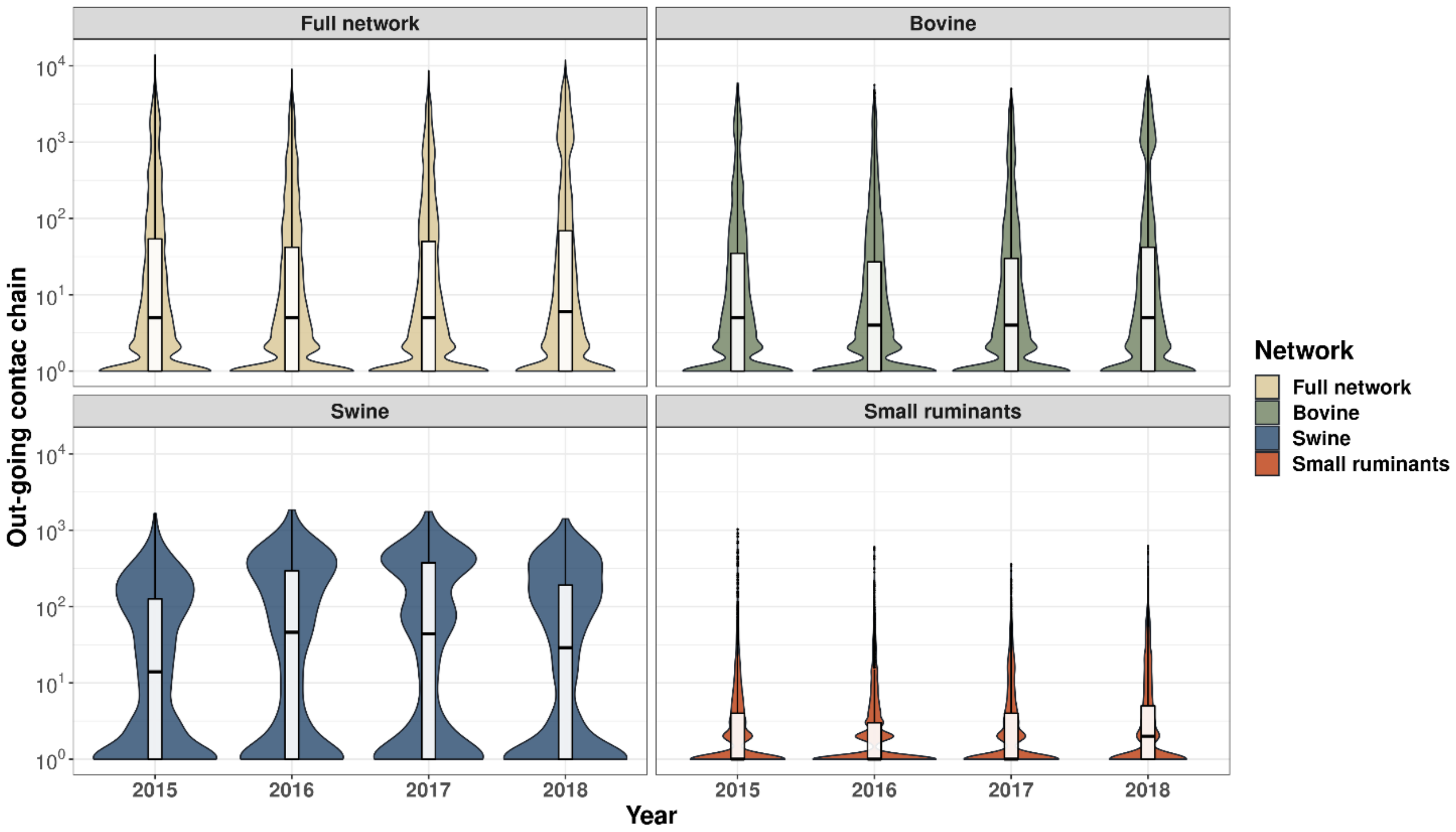

**Figure. 5** Boxplots and violin plots of the distribution of the yearly out-going contact chains by host and for the full network.

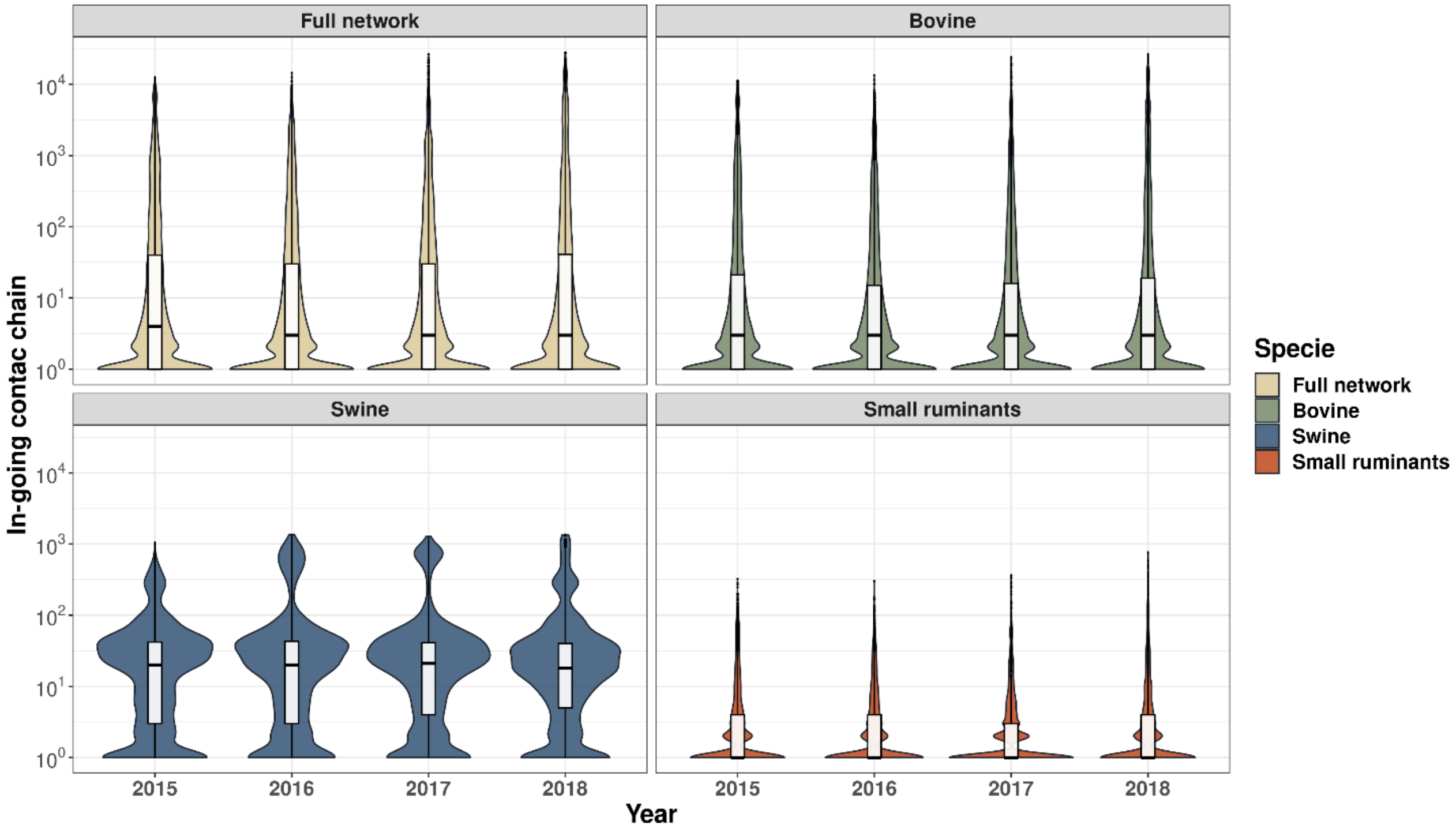

**Figure 6.** Boxplots and violin plots of the distribution of the yearly in-going contact chains by host and for the full network.

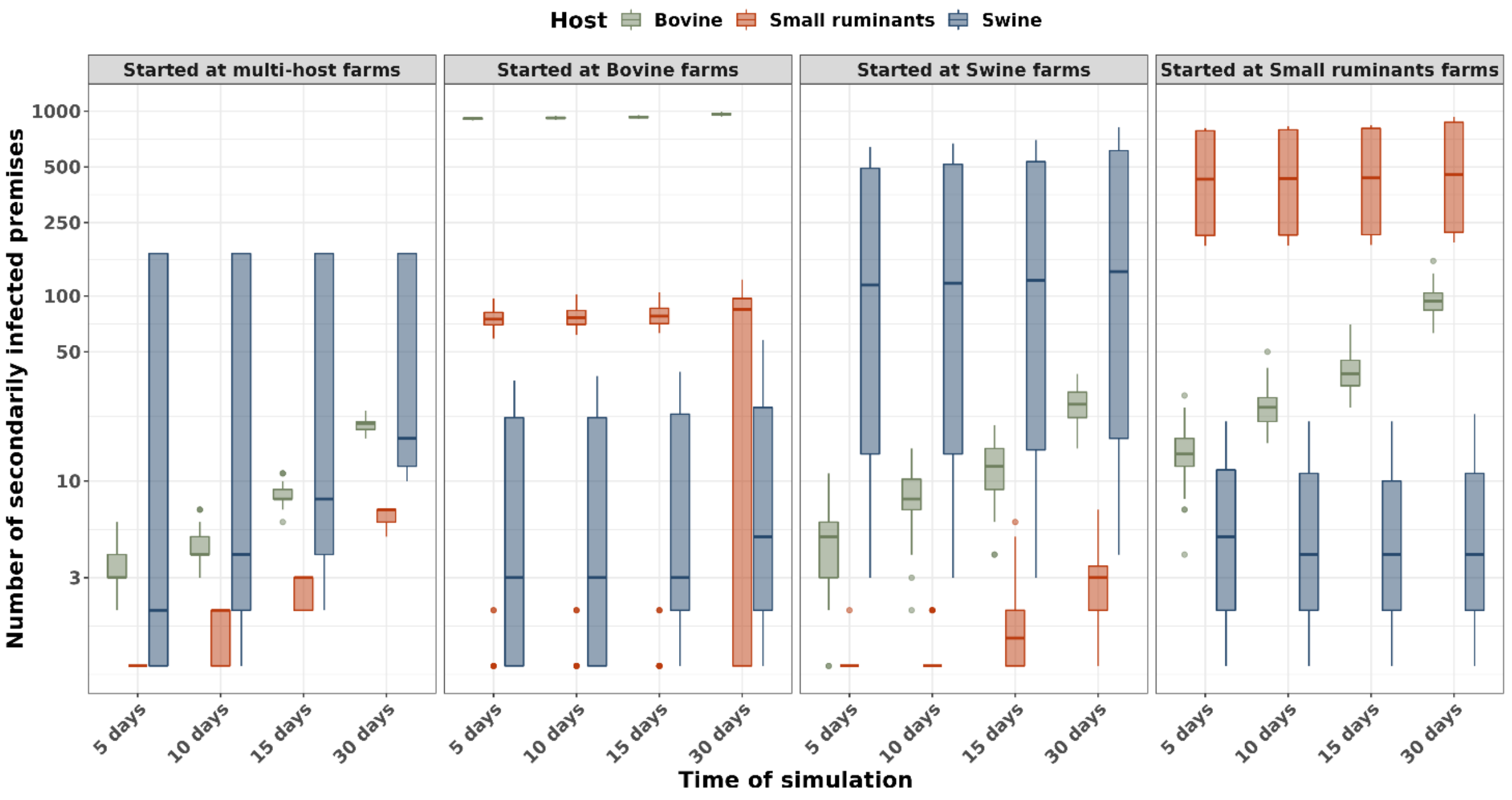

**Figure 7. Number of secondarily infected farms over 30 days post-seed infection.** Each title plot represents the host in which the infection was seeded. The result is represented in boxplots where the y-axis is in log10 scale and represents the number of infected farms prevalence generated through 100 simulations of each host for the first 5, 10, 15, and 30 days post introduction.

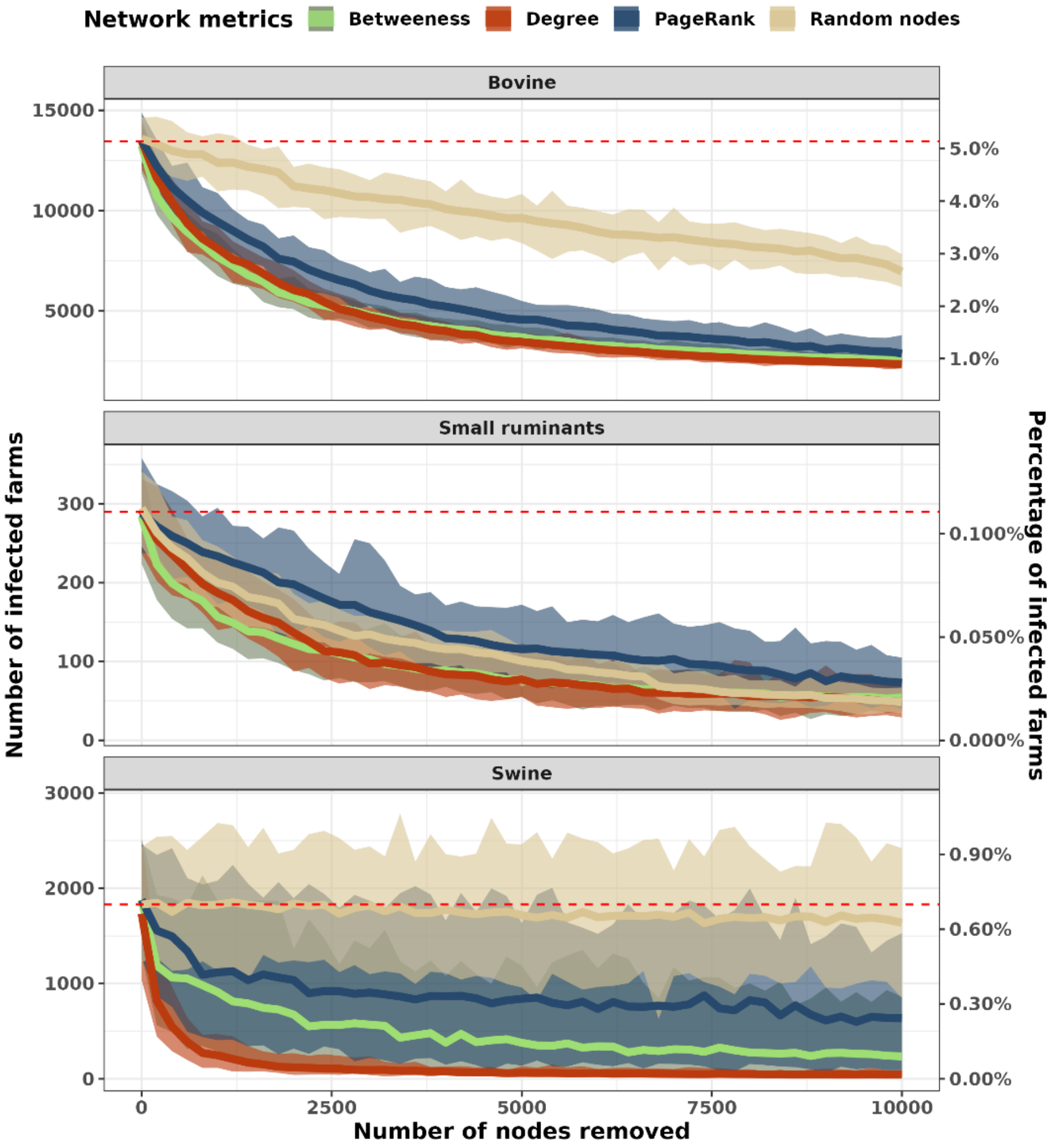

**Figure 8. Characterization of disease dynamics in terms of secondarily infected farms for each host species** The Simulations assumed a farm prevalence of 10% and simulation at day zero started with 1000 infected farms. Each color shaded area represents the node removal results for each network metric and the solid line represents the means of these results. The y-left axis represents the number of infected farms and the y-right axis represents the percentage of infected farms in the total population. The x-axis represents the number of farms removed from the network and the colors represent network metrics calculated here from the full network. The red dashed line represents the mean of infected farms when no nodes were removed from the network (without control scenario).

## Declarations

### Ethics approval and consent to participate

The authors confirm the ethical policies of the journal, as noted on the journal's author guidelines page. Since this work did not involve animal sampling nor questionnaire data collection by the researchers, there was no need for ethics permits.

**Consent for publication**

All authors discussed the results and critically reviewed the manuscript to approve for publication.

**Availability of data and materials**

The data that support the findings of this study are not publicly available and are protected by confidential agreements, therefore, are not available. Spread disease simulation and node removal code are available in the repository: https://github.com/machado-lab/Multiple-species-animal-movements-network-properties-disease-dynamic-and-the-impact-of-targeted

**Competing interests**

All authors confirm that there are no conflicts of interest to declare

**Funding**

Fundo de desenvolvimento e defesa sanitária animal (FUNDESA-RS) Award number 2021-1318. This funding body did not play a role in the design, analysis, and reporting of the study.

**Authors' contributions**

NCC and GM conceived the study. NCC and GM participated in the design of the study. FPNL
Coordinated the movement data collection. NCC conducted data processing, cleaning, designed the model, and simulated scenarios with the assistance of GM. NCC and GM designed the computational analysis. NCC conducted the formal coding. NCC, ALS, and GM wrote and edited the manuscript. All authors discussed the results and critically reviewed the manuscript. GM secured the funding.

**Acknowledgments**

The authors would like to acknowledge funding support from Fundo de Desenvolvimento e Defesa sanitária animal (FUNDESA-RS).

**Table 1.** Description of network analysis terminology and metrics

| **Parameter** | **Definition** | **Referenc e** |
|---|---|---|
| Nodes | Premises or slaughterhouses. | [57] |
| Edge | The link between two nodes. | |
| Degree | This is a node-level metric where we count the number of unique contacts to and from a specific node. When the directionality of the animal movement is considered, the in-going and out-going contacts are defined: out-degree is the number of contacts originating from a specific premise, and in-degree is the number of contacts coming into a specific premise. | |
| Movements | The number of animal movements. | |
| Diameter | The longest geodesic distance between any pair of nodes using the shortest possible walk from one node to another considering the direction of the edges | [57] |
| PageRank | Google PageRank measure. A link analysis algorithm that produces a ranking based on the importance for all nodes in a network with a range of values between zero and one. The PageRank calculation considers the in-degree of a given premise and the in-degree of its neighbors. | [58] |
| Betweenness | This is a node-level network metric where the extent to which a node lies on paths connecting other pairs of nodes, defined by the number of geodesics (shortest paths) going through a node. | [57] |
| Clustering coefficient | Measures the degree to which nodes in a network tend to cluster together (i.e., if A → B and B → C, what is the probability that A → C), with a range of values between zero and one. Here, we implemented the global cluster coefficient where the number of closed triplets (or 3 x triangles) in the network over the total number of triplets (both open and closed). | [57] |
| Giant weakly connected component (GWCC) | The proportion of nodes that are connected in the largest component when directionality of movement is ignored. | [57] |
| Giant strongly connected component (GSCC). | The proportion of the nodes that are connected in the largest component when directionality of movement is considered. | [57] |
| Centralization | Is a general method for calculating a graph-level centrality score based on a node-level centrality measure. The formula for this is<br>$$C(G)=sum( max(c(w), w) - c(v),v),$$<br>where c(v) is the centrality of node v. normalized by dividing by the maximum theoretical score for a graph and essentially quantifies the extent to which the network is structured around a minority of nodes, and is quantified as the summed deviation between the maximum value recorded and the values recorded for all other nodes. Values range from 0-1, with higher values indicating more extreme centralization, illustrating a relatively | [22, 57] |

reliance or concentration of off- and onto-farms shipments from/to a nodal farm at the macro-level of the entire network.

**Table 2.** Model outputs for the spread of disease and network-based surveillance model scenarios

| Model-scenario | Description | Model output |
|---|---|---|
| A | The infection was seeded at bovine farms. | Distribution of infected farms for each host species in the first 30 days. |
| B | The infection was seeded at small ruminant farms. | |
| C | The infection was seeded at swine farms. | |
| D | The infection was seeded at multi-host farms. | |
| E | The infection was seeded at random farms considering the full network (All hosts included) | Average farm prevalence of the stochastic simulations by month, discriminated by host and year. |
| Network-based targeted interventions | The model described above, where we seeded infection in each host -Bovine, Swine, Small ruminants farms. | The number of secondarily infected farms while removing the nodes. |

**Table 3.** Static network for cattle, swine, small ruminants, and for the full network.

| Host | Nodes | Edges | Mean Degree (CI95%) | Diameter | Animals | GSCC (%) | GWCC (%) | Mean Betweenness (CI95%) | Clustering Coefficient | Network centralization |
|---|---|---|---|---|---|---|---|---|---|---|
| Swine | 13,141 | 75,205 | 11.4 (10.99-11.90) | 20 | 62,292,763 | 1,885 (14.34%) | 9,788 (74.34%) | 629,319 (4,773.80 – 7,812.57) | 0.032 | 0.219 |
| Cattle | 261,892 | 840,824 | 6.4 (6.36-6.48) | 33 | 20,773,516 | 105,609 (40.33%) | 249,742 (95.36%) | 712,451 (670,851.1-754,051) | 0.025 | 0.003 |
| Small ruminant | 32,325 | 51,848 | 3.2 (3.12-3.27) | 26 | 1,763,834 | 5,301 (16.40%) | 25,420 (78.64%) | 33,394 (29,706-36,821) | 0.028 | 0.004 |
| Full network | 277,004 | 1,038,346 | 7.12 (7.05-7.18) | 30 | 90,090,619 | 12,1975 (44.03%) | 264,908 (95.63%) | 772,436 (706,188-785,513) | 0.03 | 0.003 |

**Additional files**

Additional file 1. **Supplementary material for multiple species animal movements: network properties, disease dynamics and the impact of targeted control actions manuscript.**

Material with the description of Spatial description of farms in the study area, monthly out and in going contact chains, a monthly distribution of betweenness, number of prevalence over 30 days post-seed infection, The supplementary Table 3. The mean of all simulation farms prevalence for each month by host and year starting at all random hosts.

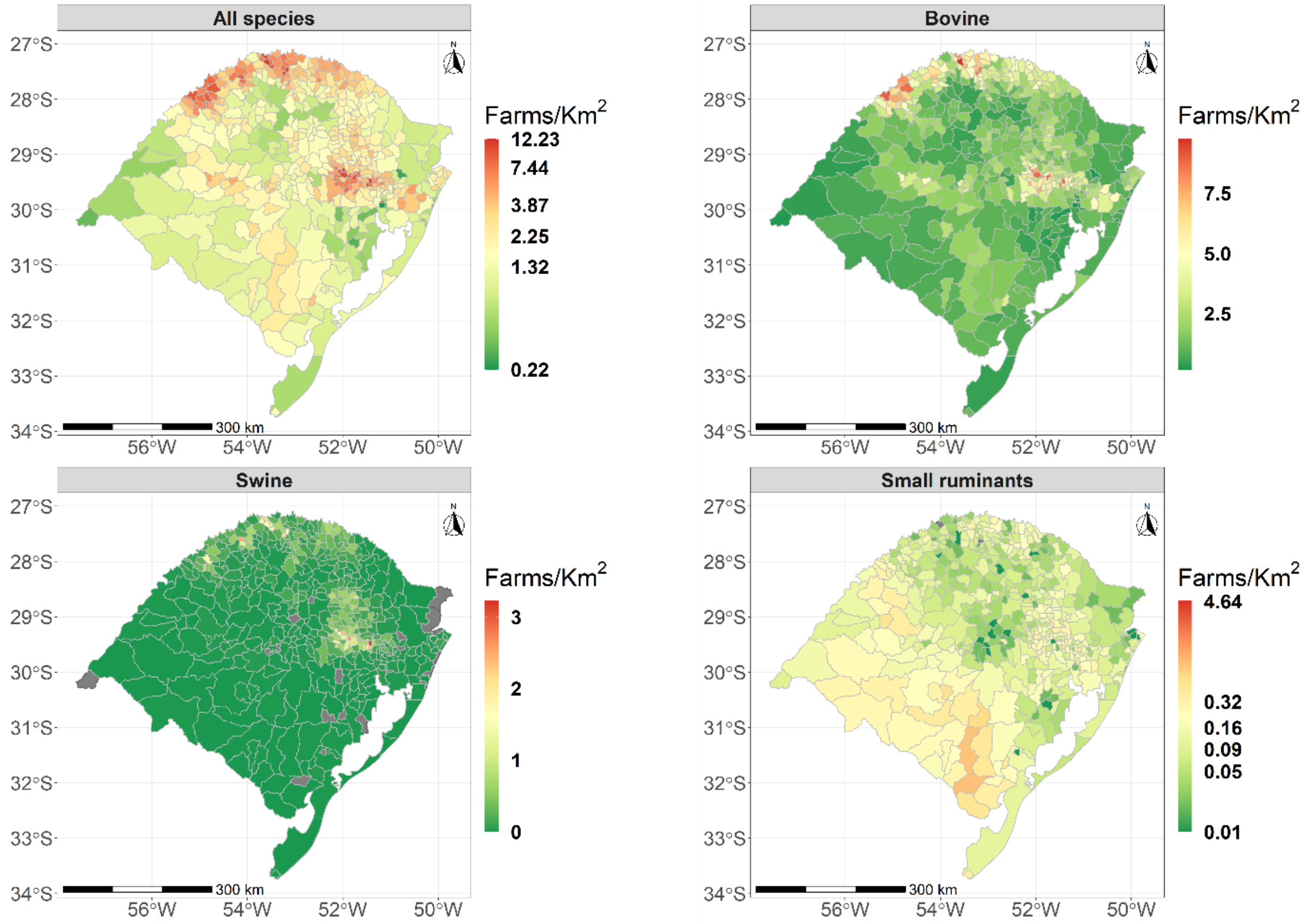

**Supplementary Figure 1**. Spatial distribution of farms by host and municipalities by Km$^2$ in the State of Rio Grande do Sul, Brazil.

**Supplementary Figure 2.** Monthly out-going contact chain distribution by host and species.

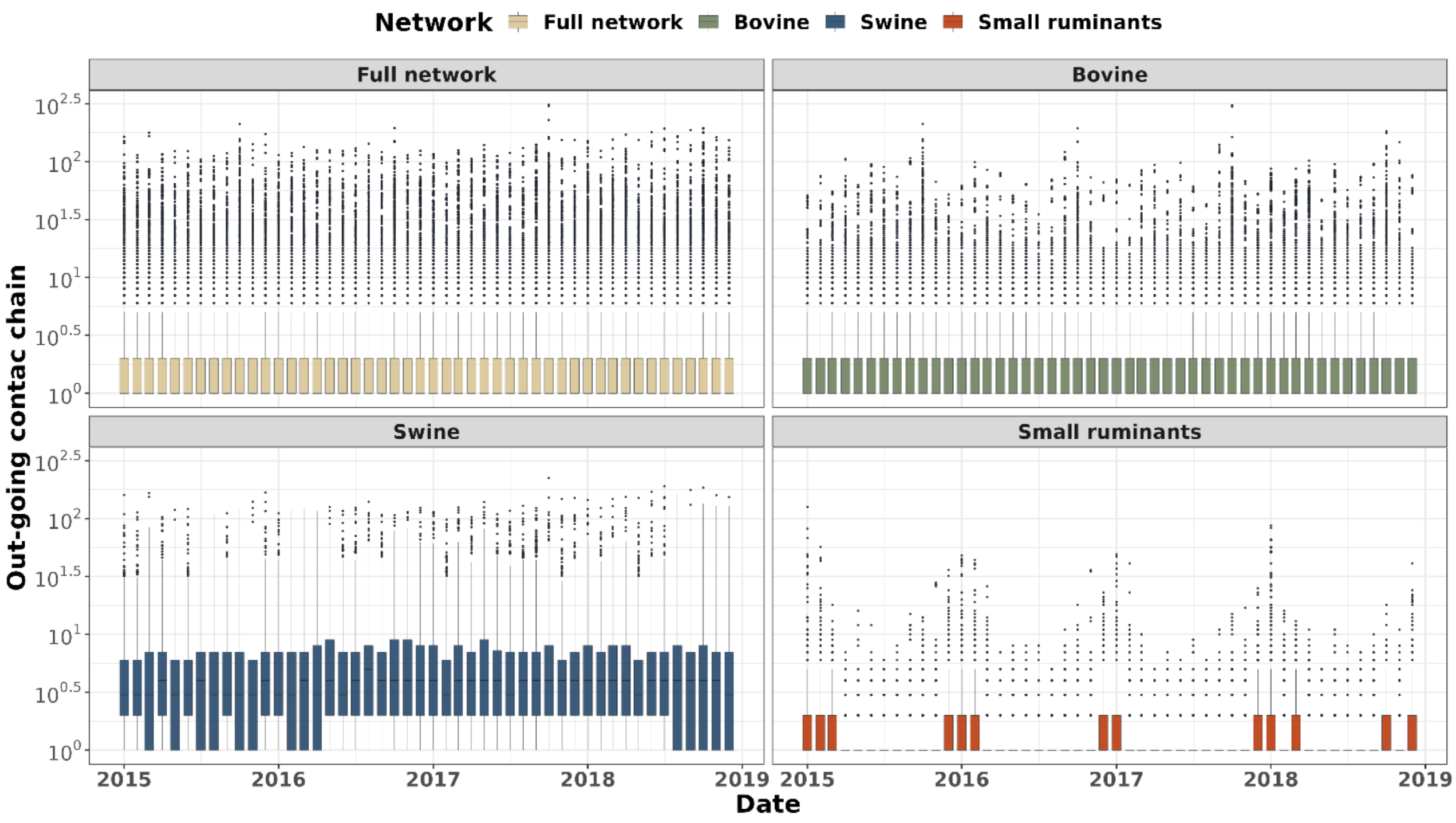

**Supplementary Figure 3.** Monthly outgoing contact chain distribution by host and species.

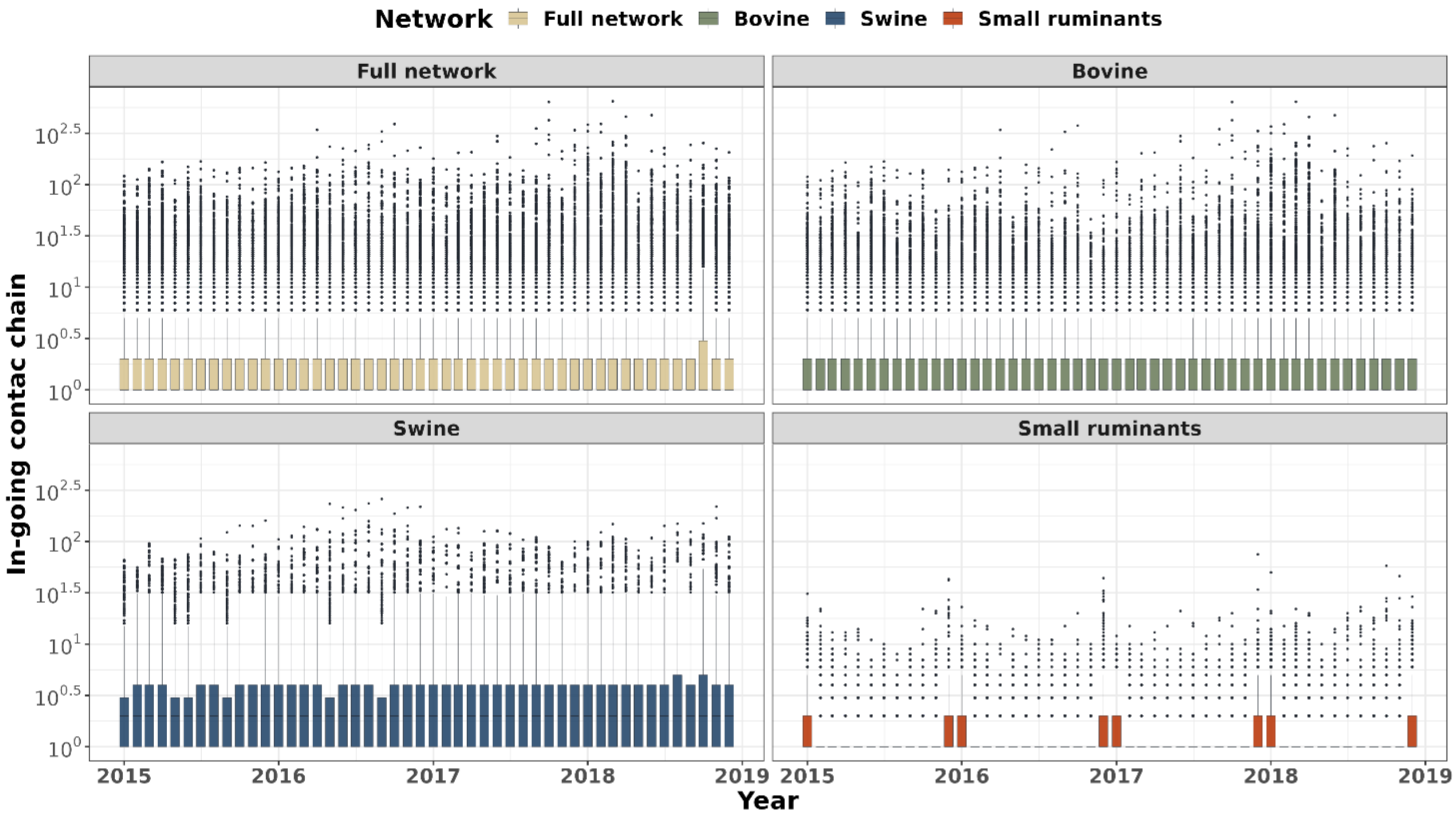

**Supplementary Figure 4.** Monthly distribution of betweenness.

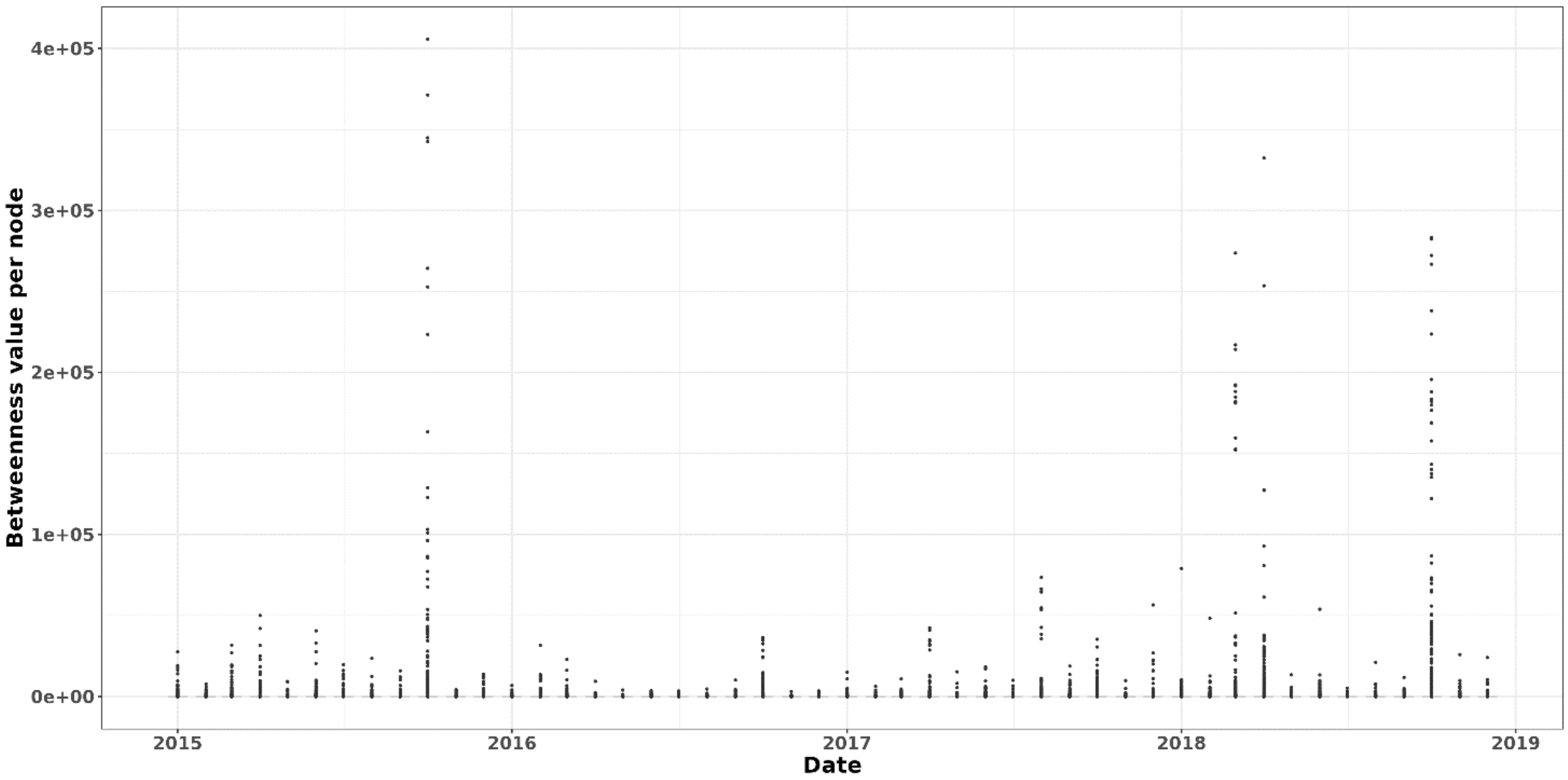

**Supplementary Figure 4. Number of prevalence over 30 days post-seed infection.** Each title plot represents the host in which the infection was seeded. The result is represented in boxplots where the y-axis is in log10 scale and represents the number of infected farms prevalence generated through 100 simulations of each host for the first 5, 10, 15, and 30 days post introduction.

**Supplementary Table 1.** The number of records and animals moved among hosts.

| Time_of_simulation | Type farm where the infection started | Affected host | Min_prevalence | Max_prevalence | Mean_prevalence | Median_prevalence | IQR_prevalence |
|---|---|---|---|---|---|---|---|
| 5 days | Started at Bovine farms | Bovine | 0.341 | 0.357 | 0.349 | 0.349 | 0.005 |
| 5 days | Started at Bovine farms | Small ruminants | 0.003 | 0.298 | 0.215 | 0.231 | 0.036 |
| 5 days | Started at Bovine farms | Swine | 0.008 | 0.267 | 0.085 | 0.023 | 0.160 |
| 5 days | Started at multis-host farms | Bovine | 0.001 | 0.002 | 0.001 | 0.001 | 0.000 |
| 5 days | Started at multis-host farms | Small ruminants | 0.003 | 0.003 | 0.003 | 0.003 | 0.000 |
| 5 days | Started at multis-host farms | Swine | 0.008 | 1.294 | 0.438 | 0.015 | 1.287 |
| 5 days | Started at Small ruminants farms | Bovine | 0.002 | 0.011 | 0.006 | 0.005 | 0.002 |
| 5 days | Started at Small ruminants farms | Small ruminants | 0.575 | 2.499 | 1.533 | 1.528 | 1.758 |
| 5 days | Started at Small ruminants farms | Swine | 0.008 | 0.160 | 0.051 | 0.038 | 0.072 |
| 5 days | Started at Swine farms | Bovine | 0.000 | 0.004 | 0.002 | 0.002 | 0.001 |
| 5 days | Started at Swine farms | Small ruminants | 0.003 | 0.006 | 0.003 | 0.003 | 0.000 |
| 5 days | Started at Swine farms | Swine | 0.023 | 4.888 | 2.018 | 1.607 | 3.643 |
| 10 days | Started at Bovine farms | Bovine | 0.343 | 0.361 | 0.352 | 0.352 | 0.006 |
| 10 days | Started at Bovine farms | Small ruminants | 0.003 | 0.314 | 0.207 | 0.235 | 0.041 |
| 10 days | Started at Bovine farms | Swine | 0.008 | 0.282 | 0.084 | 0.023 | 0.160 |
| 10 days | Started at multis-host farms | Bovine | 0.001 | 0.003 | 0.002 | 0.002 | 0.000 |
| 10 days | Started at multis-host farms | Small ruminants | 0.003 | 0.006 | 0.005 | 0.006 | 0.003 |
| 10 days | Started at multis-host farms | Swine | 0.008 | 1.294 | 0.446 | 0.030 | 1.279 |
| 10 days | Started at Small ruminants farms | Bovine | 0.006 | 0.019 | 0.010 | 0.010 | 0.003 |
| 10 days | Started at Small ruminants farms | Small ruminants | 0.578 | 2.545 | 1.550 | 1.545 | 1.787 |
| 10 days | Started at Small ruminants farms | Swine | 0.008 | 0.160 | 0.049 | 0.030 | 0.069 |
| 10 days | Started at Swine farms | Bovine | 0.001 | 0.006 | 0.003 | 0.003 | 0.001 |
| 10 days | Started at Swine farms | Small ruminants | 0.003 | 0.006 | 0.004 | 0.003 | 0.000 |
| 10 days | Started at Swine farms | Swine | 0.023 | 5.102 | 2.118 | 1.679 | 3.836 |
| 15 days | Started at Bovine farms | Bovine | 0.347 | 0.364 | 0.355 | 0.355 | 0.006 |
| 15 days | Started at Bovine farms | Small ruminants | 0.003 | 0.323 | 0.206 | 0.240 | 0.046 |
| 15 days | Started at Bovine farms | Swine | 0.008 | 0.297 | 0.086 | 0.023 | 0.160 |
| 15 days | Started at multis-host farms | Bovine | 0.002 | 0.004 | 0.003 | 0.003 | 0.000 |
| 15 days | Started at multis-host farms | Small ruminants | 0.006 | 0.009 | 0.008 | 0.009 | 0.003 |
| 15 days | Started at multis-host farms | Swine | 0.015 | 1.294 | 0.460 | 0.061 | 1.264 |
| 15 days | Started at Small ruminants farms | Bovine | 0.010 | 0.027 | 0.015 | 0.015 | 0.005 |
| 15 days | Started at Small ruminants farms | Small ruminants | 0.581 | 2.591 | 1.574 | 1.561 | 1.823 |
| 15 days | Started at Small ruminants farms | Swine | 0.008 | 0.160 | 0.046 | 0.030 | 0.061 |
| 15 days | Started at Swine farms | Bovine | 0.002 | 0.008 | 0.005 | 0.005 | 0.002 |
| 15 days | Started at Swine farms | Small ruminants | 0.003 | 0.018 | 0.005 | 0.005 | 0.003 |
| 15 days | Started at Swine farms | Swine | 0.023 | 5.315 | 2.190 | 1.744 | 3.963 |
| 30 days | Started at Bovine farms | Bovine | 0.358 | 0.380 | 0.369 | 0.369 | 0.008 |
| 30 days | Started at Bovine farms | Small ruminants | 0.003 | 0.378 | 0.190 | 0.261 | 0.295 |
| 30 days | Started at Bovine farms | Swine | 0.008 | 0.442 | 0.100 | 0.038 | 0.175 |
| 30 days | Started at multis-host farms | Bovine | 0.006 | 0.009 | 0.008 | 0.008 | 0.001 |
| 30 days | Started at multis-host farms | Small ruminants | 0.015 | 0.022 | 0.020 | 0.022 | 0.003 |
| 30 days | Started at multis-host farms | Swine | 0.076 | 1.294 | 0.503 | 0.129 | 1.203 |
| 30 days | Started at Small ruminants farms | Bovine | 0.024 | 0.059 | 0.036 | 0.036 | 0.008 |
| 30 days | Started at Small ruminants farms | Small ruminants | 0.599 | 2.859 | 1.686 | 1.648 | 2.000 |
| 30 days | Started at Small ruminants farms | Swine | 0.008 | 0.175 | 0.048 | 0.030 | 0.069 |
| 30 days | Started at Swine farms | Bovine | 0.006 | 0.015 | 0.010 | 0.010 | 0.003 |
| 30 days | Started at Swine farms | Small ruminants | 0.003 | 0.022 | 0.009 | 0.009 | 0.005 |
| 30 days | Started at Swine farms | Swine | 0.030 | 6.221 | 2.524 | 2.037 | 4.550 |

**Supplementary Figure 2.** Descriptive statistics for the farm prevalence in the 5, 10, 15, 30, days of spread disease simulation.

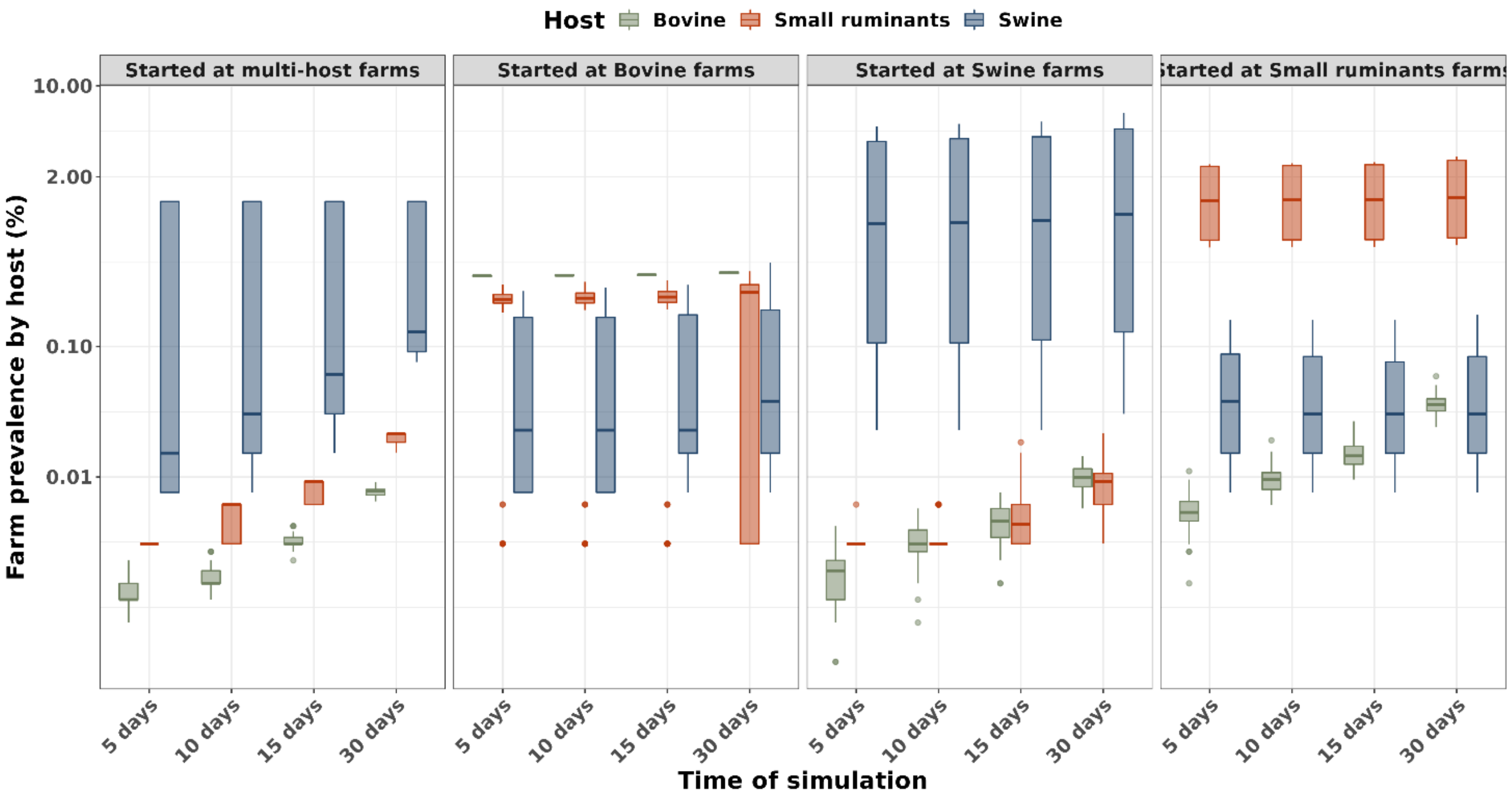
Host
Bovine
Small ruminants
Swine
Started at multi-host farms
Started at Bovine farms
Started at Swine farms
Started at Small ruminants farms
10.00
2.00
0.10
0.01
Farm prevalence by host (%)
5 days
10 days
15 days
30 days
Time of simulation